\documentclass[apjl]{emulateapj}
\usepackage{txfonts}
\usepackage[dvipsnames,usenames]{xcolor}
\usepackage{natbib}

\newcommand{\code}[1]{\texttt{#1}}
\newcommand*{\maxi}{MAXI~J0556-332}
\newcommand*{\xte}{XTE~J1701-462}
\newcommand*{\qimp}{$Q_{\rm imp}$}                   
\newcommand*{\MeV}{\mathrm{MeV}}                   
\newcommand*{\grampercc}{\mathrm{g\,cm^{-3}}}
\newcommand*{\K}{\mathrm{K}}
\newcommand*{\emcee}{\code{emcee}}
\newcommand*{\mesa}{\code{MESA}}
\newcommand*{\dStar}{\code{dStar}}

\begin{document}
\title{A Strong Shallow Heat Source in the Accreting Neutron Star MAXI~J0556-332}
\author{Alex Deibel\altaffilmark{1,2$^{\star}$}, Andrew Cumming\altaffilmark{2,3}, Edward F. Brown\altaffilmark{1,2,4}, and Dany Page\altaffilmark{2,5}}
\affil{
\altaffilmark{1}{Department of Physics and Astronomy, Michigan
State University, East Lansing, MI 48824, USA} \\
\altaffilmark{2}{The Joint Institute for Nuclear Astrophysics - Center for the Evolution of the Elements, Michigan State University, East Lansing, MI 48824, USA} \\
\altaffilmark{3}{Department of Physics, McGill University, 3600 rue University, Montreal, QC, H3A 2T8, Canada} \\
\altaffilmark{4}{National Superconducting Cyclotron Laboratory, Michigan State University,
East Lansing, MI 48824, USA} \\
\altaffilmark{5}{Instituto de Astronom\'{i}a, Universidad Nacional Aut\'{o}noma de M\'{e}xico, Mexico D.F. 04510, Mexico}
}
\shorttitle{Strong Shallow Heating in \maxi}
\shortauthors{Deibel et al.}

\email{$^{\star}$deibelal@msu.edu}

\begin{abstract}
An accretion outburst in an X-ray transient deposits material onto the neutron star primary; this accumulation of matter induces reactions in the neutron star's crust. During the accretion outburst these reactions heat the crust out of thermal equilibrium with the core. When accretion halts, the crust cools to its long-term equilibrium temperature on observable timescales. Here we examine the accreting neutron star transient \maxi, which is the hottest transient, at the start of quiescence, observed to date. Models of the quiescent light curve require a large deposition of heat in the shallow outer crust from an unknown source. The additional heat injected is $\approx 4\textrm{--}10\,\MeV$ per accreted nucleon;  when the observed decline in accretion rate at the end of the outburst is accounted for, the required heating increases to $\approx 6\textrm{--}16\,\MeV$. This shallow heating is still required to fit the lightcurve even after taking into account a second accretion episode, uncertainties in distance, and different surface gravities. The amount of shallow heating is larger than that inferred for other neutron star transients and is larger than can be supplied by nuclear reactions or compositionally driven convection; but it is consistent with stored mechanical energy in the accretion disk. The high crust temperature ($T_b \gtrsim 10^{9} \, {\rm K}$) makes its cooling behavior in quiescence largely independent of the crust composition and envelope properties, so that future observations will probe the gravity of the source. Fits to the lightcurve disfavor the presence of Urca cooling pairs in the crust.
\end{abstract}


\section{Introduction}
\label{sec:intro}

With the fading of the neutron star KS~1731-260 into quiescence in 2001, it was quickly realized that long-term monitoring of the quiescent light curve of quasi-persistent transients offers an opportunity to study the properties of dense matter \citep{ushomirsky2001,wijnands2001,rutledge2002}. During accretion onto a quasi-persistent transient, the continual compression of the crust induces electron capture, neutron emission, and pycnonuclear reactions that release $1$--$2$ $\MeV$ per accreted nucleon \citep{haensel1990,haensel2008}, heating the crust out of thermal equilibrium with the core. 
When accretion halts, the crust thermally relaxes back to the core temperature on observable timescales \citep{rutledge2002}. This relaxation has been observed in long-term monitoring of several sources, including KS 1731-260 \citep{wijnands2001, wijnands2002, cackett2010}, MXB 1659-29 \citep{wijnands2003, wijnands2004, cackett2008}, XTE~J1701-462  \citep{fridriksson2010, fridriksson2011} and EXO~0748-676 \citep{degenaar09, degenaar2014}. Theoretical models of crust relaxation successfully reproduce quiescent cooling curves with interesting constraints on crust physics, such as the thermal conductivity of the inner crust \citep{shternin2007,brown09,page2013,horowitz2015,turlione2015}.

A surprising find is that the shallow outer crust where the local density is $\lesssim {10^{10}}{\,\grampercc}$ must be substantially heated with respect to the deeper neutron star crust in order to explain the temperatures observed in the first months of relaxation. In models of KS 1731-260 and MXB 1659-29, \cite{brown09} required an additional heat source of $\approx 1 \, \MeV$ at these depths. The physical source of this shallow heating is as yet unknown, but the need for it has also been inferred in studies of thermonuclear burning on accreting neutron stars \citep{cumming2006,keek2008,intzand2012,zamfir2014}. Motivated by these findings, there have been extensive observational efforts to follow the light curves of other quasi-persistent transients at early times to catch the thermal relaxation of these shallow outer layers, e.g.~IGR~J17480-2446 \citep{degenaar2013}.

In this paper, we present fits to the light curve of the transient source \maxi \ (hereafter MAXI) \citep{matsumura2011} which place the most demanding requirements yet on the shallow heat source. MAXI shows many similarities to the class of low mass X-ray binaries known as Z-sources \citep{homan2011, sugizaki2013}, which implies that the compact object is a neutron star accreting at near the Eddington rate and is at a large distance $\sim 46 \pm 15 \, {\rm kpc}$, as determined from the observed flux \citep{homan14}. MAXI is by far the hottest quiescent neutron star in this class observed to date. An exponential fit to the declining temperature gives a large drop in temperature on a timescale comparable to the shortest timescales observed in other sources. The cooling curve of MAXI is reproduced naturally by crust models if a shallow heat source of $Q_{\rm shallow} \approx$ $4$--$10$ $\MeV$ is included. The high crust temperature means that the physical conditions in the crust are in a different regime than the cooler sources such as KS 1731-260 and MXB 1659-29, making MAXI a particularly interesting test of the crust cooling scenario. The large temperature also allows us to constrain whether Urca cooling is operating in the outer crust where it is expected to balance crustal heating during accretion \citep{schatz2014} .

In Section~\ref{sec:models}, we outline the crust thermal relaxation model and the fit to the MAXI light curve. In Section~\ref{sec:reflare}, we explore the sensitivity of our light curve fits to the assumptions, such as the distance to the source, accretion rate variations during outburst, and the choice of gravity. In Section~\ref{sec:Urca}, we use the MAXI light curve to constrain the presence of Urca cooling pairs. We discuss our results in Section~\ref{sec:discussion}, in particular the implications of such a strong heat source for models of the shallow heating mechanism.

\begin{figure}
\begin{center}
\includegraphics[width=1.0\columnwidth]{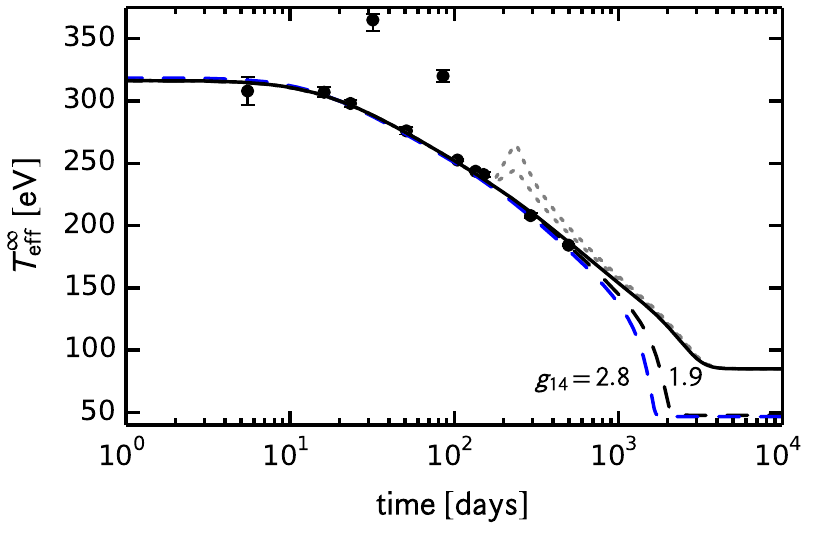}
\end{center}
\vspace{-0.4cm}
\caption{Model fit to the quiescent light curve of \maxi . The solid black curve corresponds to a model with $M=1.5\ M_\odot$, $R=11\ {\rm km}$, $Q_{\rm shallow} = 6.0\, \MeV$, and $T_{\rm core}=10^8 \, {\rm K}$; the dashed black curve is for the same model with $T_{\rm core}=3\times 10^7 \, {\rm K}$. The black dotted curves are light curves with a reheating event $\approx 170$ days into quiescence for $Q_{\rm shallow} = 6.0\, \MeV$ (upper curve) and $Q_{\rm shallow} = 3.0\, \MeV$ (lower curve). The blue dashed curve is for a $M=2.1\ M_\odot$, $R=12\ {\rm km}$ neutron star fit to the observations by changing the shallow heating depth and strength. the data above the light curve are contamination from residual accretion. Note that $T_{\mathrm{eff}}^{\infty} \propto g^{1/4}/(1+z)$ which leads to different observed core temperatures for different gravities.}
\label{fig:lightcurve_dStar}
\end{figure}

\section{Crust thermal relaxation models of the MAXI~J0556-332 light curve}
\label{sec:models}

We solve the thermal evolution of the neutron star crust numerically by evolving the thermal diffusion equation. To provide a check on our results, we do this with two different numerical implementations. The first is the open-source code \dStar\footnote{\code{https://github.com/nworbde/dStar}} which solves the fully general relativistic heat diffusion equation for the crust using the method of lines, implemented using stiff ODE solvers in the \mesa\ numerical library \citep{paxton2011, paxton2013}.  The second code \code{crustcool}\footnote{\code{https://github.com/andrewcumming/crustcool}}, assumes a constant Newtonian gravity and applies a global redshift to the observer frame. This is a good approximation because the crust is thin and is more efficient computationally for fitting purposes. To perform Markov chain Monte Carlo fits, we have coupled the \code{crustcool} calculations to the \emcee\ code \citep{foreman2013}. The microphysics input in both codes (equation of state, thermal conductivity, superfluid critical temperatures, and neutrino emissivities) is similar and follows \citet{brown09}.


The temperature $T_b$ at the top of the computational grid (typically taken at a column depth $y=10^{10}\ {\rm g\ cm^{-2}}$) is mapped to the photosphere temperature $T_{\rm eff}$ using a separately computed set of envelope models with a helium top-layer and iron bottom-layer (following \citealt{brown02}). At the temperatures observed for MAXI, the $T_{\rm eff}\textrm{--}T_b$ relation is insensitive to the helium mass in the envelope. Whereas \citet{brown09} held $T_b$ fixed during accretion to simulate the effect of a shallow heat source, we instead include the heat source directly and allow $T_b$ to evolve as accretion proceeds. The shallow heat source is uniformly distributed in $\log y$ centered on a value $y_h = 6.5\times 10^{13}\ {\rm g\ cm^{-2}} $ ($\rho\approx 1.2\times 10^{10}\ {\rm g\ cm^{-3}}$) and ranging from $y_h/3$ to $y_h\times 3$. The strength of shallow heating is assumed to vary proportionally with the accretion rate.  

\begin{figure}
\begin{center}
\includegraphics[width=1.0\columnwidth]{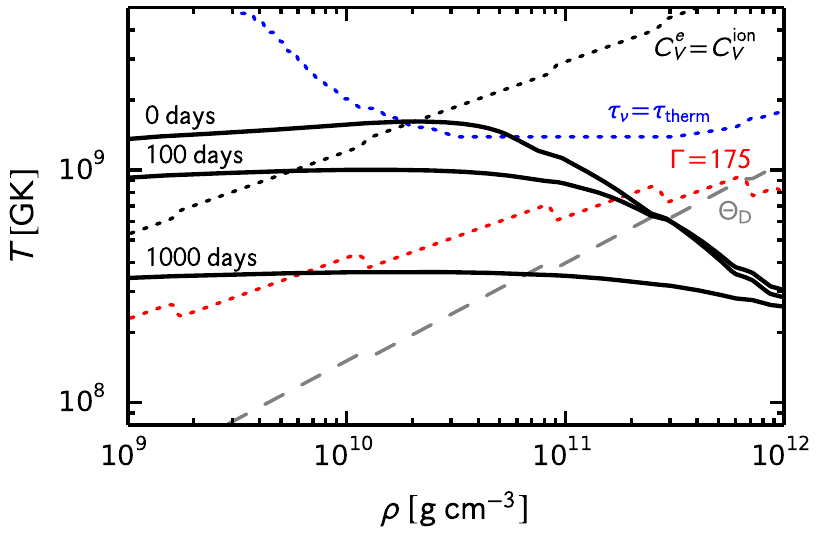}
\end{center}
\vspace{-0.4cm}
\caption{Solid black curves indicate the evolution of the crust temperature during quiescence for the $M=1.5\ M_\odot$ and $R=11\ {\rm km}$ model, shown in Figure \ref{fig:lightcurve_dStar}. The red dotted curve is the melting line of the crust ($\Gamma = 175$) for the crust composition in \citet{haensel1990}, the black dotted curve is the transition from an electron-dominated heat capacity to an ion-dominated heat capacity ($C_V^e = C_V^{\rm ion}$), and the blue dotted curve is where the local neutrino cooling time is equal to the thermal diffusion time ($\tau_{\nu} = \tau_{\rm therm}$). The gray dashed curve shows the lattice Debye temperature $\Theta_{\rm D}$; when $T \ll\Theta_{\rm D}$ electron-impurity scattering influences the thermal conductivity.}
\label{f.t_profile}
\end{figure}

To fit the cooling light curve, we assume that a $M=1.5\ M_\odot$ and $R=11\ {\rm km}$ neutron star accreted at the local Eddington rate $\dot m = \dot m_{\rm Edd} \equiv 8.8 \times 10^{4}\ {\rm g \ cm^{-2} \ s^{-1}}$ for 16 months, matching the duration of the MAXI outburst \citep{homan14}, before cooling began. We find that the subsequent cooling of the crust naturally reproduces the shape of the light curve if we include a strong shallow heat source. The solid curve in Figure \ref{fig:lightcurve_dStar} shows a model with a $Q_{\rm shallow} = 6.0\, \MeV$ heat source (the other curves will be discussed in Section~\ref{sec:reflare}). The temperature at the top of the crust reaches $T_b \simeq 2 \times 10^{9}\, {\rm K}$ by the end of outburst, as shown in Figure~\ref{f.t_profile}. At the high temperatures found in the crust of MAXI the electron thermal conductivity is controlled by electron-ion and electron-phonon scattering in the liquid and solid phase, respectively. It is only when $T\ll \Theta_{\rm D}$, where $\Theta_{\rm D}$ is the lattice Debye temperature, that electron-impurity scattering influences the thermal conductivity. In MAXI, the crust temperature is always well above $\Theta_{\rm D}$ and electron-impurity scattering plays no role in the thermal conductivity. Therefore, we set the impurity parameter, which determines the electron-impurity scattering contribution to the thermal conductivity, to be $Q_{\rm imp}=1$.



A few analytic estimates help outline the location and strength of shallow heating needed to explain the light curve. The break in the light curve at $\approx 10$--$20$ days into quiescence occurs when the inward propagating cooling front reaches the shallow heating depth (i.e., the peak of the crust temperature profile). The time for the region of the crust with mass density $\rho$ to cool is its thermal time, $\tau_{\rm therm} = ({1}/{4}) [ \int_{z} (\rho C_V/K)^{1/2} dz' ]^2$ \citep{henyey69}, where $C_V$ is the specific heat and $K$ is the thermal conductivity given by $K=(\pi^2/3)(n_e c^2 k_{\mathrm{B}}^2T/E_{\mathrm{F}}\nu)$ with $n_e$ the electron density, $E_{\rm F}$ the electron fermi energy, and $\nu$ the electron collision frequency. \citet{brown09} showed that $\tau_{\rm therm}\propto \rho$, independent of temperature, when the heat capacity is dominated by the ions in the solid lattice and electron-phonon scattering dominates the thermal conductivity (see Equation~(9) of \citealt{brown09}). The physical conditions appropriate for the early phase of the MAXI cooling curve are quite different because the temperatures reached during outburst are high enough to melt the crust at densities $\rho\lesssim 2 \times 10^{11} \, {\rm g \ cm^{-3}}$. The appropriate choices for $C_V$ and $\nu$ are then the electron heat capacity $C_{V}^{e} =\pi^2(Z/A)(k_{\mathrm{B}}/m_p)(k_{\mathrm{B}}T/E_{\mathrm{F}})$ and electron-ion scattering $\nu=\nu_{ei} = 4e^4(E_{\mathrm{F}}/c^2) Z\Lambda_{ei}/3\pi\hbar^3$ \citep{yakovlev80}, where the electron Fermi energy is $E_{\mathrm{F}}=3.7\ {\rm MeV}\ (\rho_9Y_e/0.4)^{1/3}$ and $\Lambda_{ei}\approx 1$ is the Coulomb logarithm. The resulting thermal time is 
\begin{equation}\label{eq:ttherm}
\tau_{\rm therm, liquid}^{\infty} \approx 1.2\, {\rm days} \ \rho_{9} \ \left(\frac{g_{14}}{2}\right)^{-2} \left({Y_e\over 0.4}\right)^3\left({Z\over 34}\right) \left({1+z\over 1.24}\right) \ ,
\end{equation}
where we introduce the parameters $g_{14} \equiv g/10^{14} \, {\rm cm \ s^{-2}}$ and $g=(1+z)GM/R^2$, where $1+z = (1-2GM/(Rc^2))^{-1/2}$ redshifts to an observer frame at infinity. Interestingly, the temperature independent scaling $\tau_{\rm therm}\propto\rho$ still holds in this case\footnote{Since $C_{V}^{e}/C_{V}^{\mathrm{ion}}=0.82(Z/34)^{8/3}(T/T_{\rm melt})$ (see Fig.\ \ref{f.t_profile}), the outer crust is either mostly liquid with a heat capacity dominated by electrons, or mostly solid with a heat capacity dominated by ions. In general, therefore, $\tau_{\rm therm}\propto \rho$ and is independent of temperature.}.
The shallow heat source needs to be located at a density $\sim 10^{10} \ {\rm g\ cm^{-3}}$ for the light curve to break at tens of days, consistent with the break in the MAXI light curve. Rearranging Equation~(\ref{eq:ttherm}), and using the fact that the electrons are degenerate, gives the pressure at the heating depth
\begin{equation}\label{eq:yheat}
P_{h,28} \approx 1.5 \ \left({t\over 20\ {\rm days}}\right)^{4/3} \ \left(\frac{g_{14}}{2}\right)^{8/3} \left(\frac{Y_e}{0.4}\right)^{-8/3} \left(\frac{34}{Z} \right)^{4/3} \ ,
\end{equation}
in units of $P_{28} \equiv 10^{28} \ {\rm ergs \ cm^{-3}}$, where $t$ is the time of the break in the light curve tens of days into quiescence. 


To estimate the strength of the shallow heating, we first use the $T_{\rm eff}$--$T_{b}$ relation for an iron envelope from \cite{gpe82}, which is approximately correct at these high temperatures, to obtain the temperature at the heating depth
\begin{equation} \label{e.teff_tb}
T_{b} = 1.2 \times 10^9\ {\rm K}\ \left({T_{{\rm eff},\infty}\over 300\ {\rm eV}}\right)^{1.82}\left({g_{14}\over 2}\right)^{-9/20} \ .
\end{equation}
Because most of the heat is transported inward, we can estimate the heating required to maintain this temperature using the inward flux $F \approx KT/H$, where $H=P/\rho g=1.5\times 10^{3} \ {\rm cm}\ y_{12}^{1/4} (g_{14}/2)^{-3/4} (Y_e/0.4)$ is the pressure scale height and the column depth is $y \approx P/g$. The heating required to maintain the inward flux is

\begin{equation}\label{eq:Qin}
Q_{\rm in} = 3.4\ \MeV \, \mathrm{u}^{-1} \ P_{28}^{1/4}\left({T_{{\rm eff},\infty}\over 300\ {\rm eV}}\right)^{1.82}\left({\dot m\over \dot m_{\mathrm{Edd}}}\right)^{-1}\left({g_{14}\over 2}\right)^{11/20} ,
\end{equation}
which agrees well with the values we find in the numerical model. At the relevant temperatures $\gtrsim 10^9$ K, given by Equation~(\ref{e.teff_tb}), neutrino cooling can be important, making Equation~(\ref{eq:Qin}) a lower limit on the heating strength. To illustrate this, we show in Figure~\ref{f.t_profile} the temperature where the local neutrino cooling timescale is comparable to the thermal time of the layer.

\begin{figure}
\begin{center}
\includegraphics[width=1.05\columnwidth]{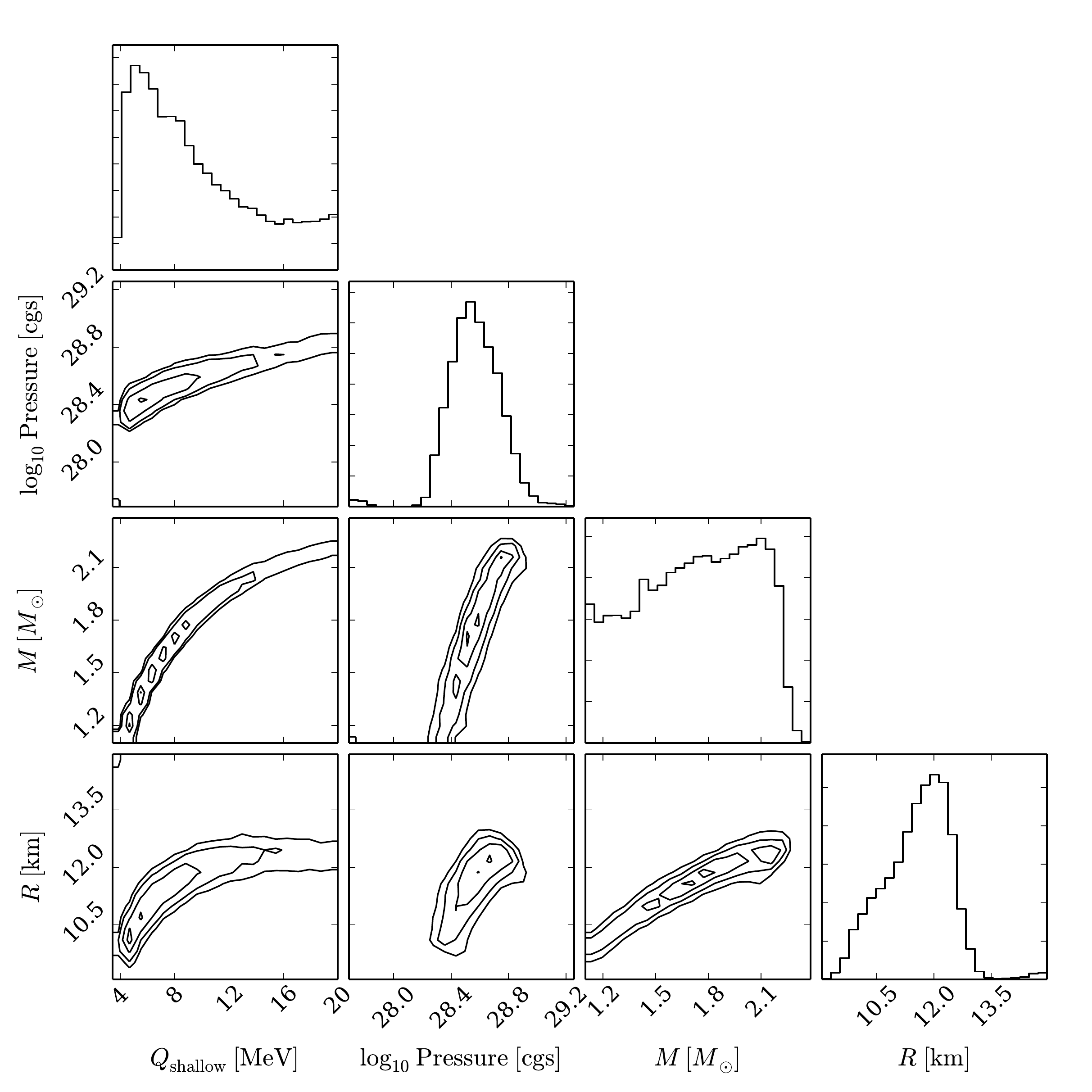}
\end{center}
\vspace{-0.4cm}
\caption{Markov chain Monte Carlo fits of \code{crustcool} to the quiescent light curve of \maxi. The contours show the isodensity surfaces of the likelihood $\mathcal{L}$, corresponding to $\sqrt{-2 \mathrm{ln}\mathcal{L}} = 0.5, \, 1, \, 1.5, \, 2$, for the neutron star mass $M$, radius $R$, pressure at the shallow heating depth $P_h$, and the shallow heating strength $Q_{\rm shallow}$. The scalings of $P_h$ and $Q_{\rm shallow}$ with $M$ and $R$ are show in Section~\ref{sec:models}.}
\label{f.mc_plot}
\end{figure}

\ \\

\section{Dependence on gravity, distance, outburst decay, and reflaring}
\label{sec:reflare}

Given that the strength of the heat source is much larger than the $1\textrm{--}2\,\MeV$ typically discussed for crust heating, we now check the sensitivity of the parameters of the shallow heat source to the various assumptions that go into the model.

The depth and strength of the heating is sensitive to the choice of neutron star mass and radius. To investigate this further, we fit the light curve using \code{emcee} and \code{crustcool} with a broad prior on $M$ and $R$ as well as a $0\textrm{--}20 \, \MeV$ prior on the shallow heating strength. The results of these models can be seen in Figure~\ref{f.mc_plot} and example light curves can be seen in Figure~\ref{fig:lightcurve_dStar} for $g_{14} = 1.9$ (black dashed curve) and $g_{14} = 2.7$ (blue dashed curve). The most probable solutions are for $1.4 \, \mathrm{M_{\odot}} \lesssim M \lesssim 2.2 \, \mathrm{M_{\odot}}$ and $10 \, \mathrm{km} \lesssim R \lesssim 12.5 \, \mathrm{km}$. In general, the higher gravity solutions are better fits to the observations, but require more shallow heating. For example, changing gravity from $1.11$ to $2.43\times 10^{14}\ {\rm cm\ s^{-2}}$ increases the heating strength required by almost a factor of 2, and increases the heating depth by almost a factor of 8. There is a tail of solutions up to $Q_{\rm shallow} \simeq 20\, \MeV$ because neutrino production at $T_{b} \gtrsim 2 \times 10^{9} \, \mathrm{K}$ consumes most of the deposited heat and prevents the crust from becoming appreciably hotter. In particular, solutions where $Q_{\rm shallow} \gtrsim 10 \, \MeV$ are allowed because of neutrino cooling in the outer crust. However, solutions where $Q_{\rm shallow} \lesssim 4\, \MeV$ are not allowed.

As suggested by \citet{homan14}, the need for a strong shallow heat source is not affected by the choice of distance to the source, which has a large uncertainty $d = 46 \pm 15 \, {\rm kpc}$. The choice of distance does alter the fit parameters for the light curve, but a strong shallow heat source is always required. For example, if we move a $g_{14}=1.11$ source closer, to $d=10\, {\rm kpc}$, $T_{\rm eff}^{\infty}$ decreases by a factor of 1.8 (see \citet{homan14}; Table 4) suggesting that less heating would be needed. Moving the source closer also decreases the inferred accretion rate, however, so if the shallow heating is proportional to accretion rate, the the heat released per accreted nucleon must actually increase to compensate. The result is that we need to increase the shallow heating from $Q_{\rm shallow}\approx 3.5 \ \MeV$ to $\approx 4.3 \ \MeV$ at a shallower depth ($y \approx 10^{13} \, {\rm g \ cm^{-2}}$) to match the light curve.


We have also investigated the effect of the outburst behavior on our conclusions. The accretion rate did not turn off instantaneously at the end of the outburst, but instead the luminosity dropped exponentially with an e-folding time of $\approx 3\ {\rm days}$ for the last $\approx 14\ {\rm days}$ of the outburst \citep{homan14}. Assuming that the accretion rate drops on a similar timescale at the end of the outburst and that the shallow heating drops proportionately to the instantaneous accretion rate changes the shape of the early quiescent light curve. Because the net crustal heating decreases proportionally with the accretion rate the star enters quiescence slightly \emph{cooler} than it would without the exponential accretion decay. That is, the thermal timescale at the heating depth (Equation~\ref{eq:ttherm}) is comparable to the turnoff time, so that as the accretion rate drops, the temperature of the cooling layer is affected by the falling accretion rate. The fact that the layer remained hot during the final phase of the outburst places a lower limit on the depth of the heat source. The net effect of the accretion decay is to increase the shallow heating required to fit the early light curve; for the $M=1.5\ M_\odot$ and $R=11\ {\rm km}$ model in Section~\ref{sec:models}, we find $Q_{\rm shallow}\approx 8.0\, \MeV$ is needed instead of $6.0\, \MeV$. 

About 170 days into quiescence, the luminosity of MAXI increased to a level similar to that observed at the end of the outburst for about 60 days \citep{homan14}. To model the reheating event, we ran a model that accretes at $\dot m \approx 0.5\, \dot m_{\rm Edd}$ for 60 days after a 170 day initial cooling phase and contains a shallow heating source. Figure~\ref{fig:lightcurve_dStar} shows the reheating light curves for two values of the shallow heating, $Q_{\rm shallow} = 6.0 \, \MeV$ (upper dotted curve) and $Q_{\rm shallow} = 3.0 \, \MeV$ (lower dotted curve). The reheated models overshoot the observations hundreds of days into quiescence and the light curve deviation lasts $\approx 500\, {\rm days}$ before returning to the cooling behavior seen prior to the reflare. We conclude that the shallow heating does not operate at the same strength during the reflare as it does during the main outburst. This implies that the shallow heating rate is not simply proportional to the accretion rate as assumed in current thermal relaxation models.


\section{Constraint on Urca cooling}
\label{sec:Urca}

\begin{figure}
\begin{center}
\includegraphics[width=1.0\columnwidth]{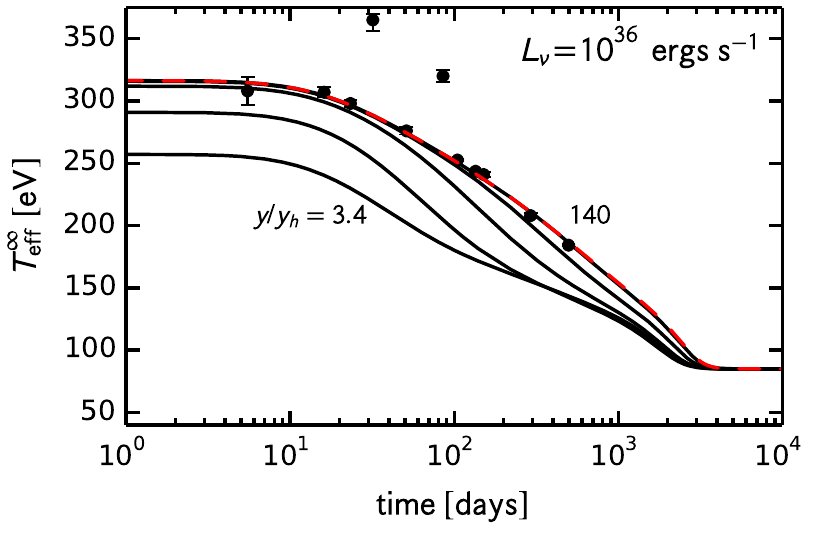}
\end{center}
\vspace{-0.4cm}
\caption{Model fit to the cooling curve from \code{dStar} with Urca shells. The crust model is for a $M=1.5\ M_\odot$, $R=11\ {\rm km}$ neutron star with $Q_{\rm shallow} = 6.0\, \MeV$. The light curve without Urca cooling is shown as a red dashed curve. Model light curves with Urca pairs, with $L_{\nu} = 10^{36} \, {\rm ergs \ s^{-1}}$, are shown as black curves. From left to right, shell depths are $y/y_h = 3.4, \, 6.7, \, 17, \, 43, \, 140$, that correspond to $\rho_{10} = 3, \, 5,\, 10, \, 20, \, 50$, respectively.}
\label{fig:Urca}
\end{figure}

The high crust temperature makes MAXI an ideal laboratory to investigate a newly discovered, potentially strong, cooling mechanism in the shallow outer crust: neutrino emission by cycles of $e^{-}$-capture and $\beta^{-}$-decay on specific pairs of nuclei \citep{schatz2014}. In a crust composed of the ashes of either X-ray bursts or superbursts, the neutrino emission from Urca pairs is sufficient to balance crust heating if the crust temperature is $\gtrsim {2}\times10^{8}\,{\K}$, well below the temperatures expected in MAXI. Confirmation of the existence of this neutrino cooling would have strong implications for our understanding of superbursts and intermediate-duration bursts \citep{intzand2005}. 

Urca cooling pairs appear in the composition over the density range $\rho \simeq 10^{10}-10^{11}\,  {\rm g \ cm^{-3}}$ and have neutrino luminosities between $L_{\nu} \simeq 10^{36}-10^{37} \, {\rm ergs \ s^{-1}}$ \citep{schatz2014}. To approximate the pair parameter space we insert Urca cooling shells at $y/y_h = 3.4, \, 6.7, \, 17, \, 43, \, 140$ (that correspond to $\rho_{10} = 3, \ 5,\ 10, \ 20, \ 50 $, respectively, where $\rho_{10} = \rho /(10^{10} \, {\rm g \ cm^{-3}})$) that have a neutrino luminosity $L_{\nu} = 10^{36} \, {\rm ergs \ s^{-1}}$, while keeping all other values in the model fixed. The light curves with Urca cooling can be seen in Figure~\ref{fig:Urca}. 

Urca cooling pairs have two effects on the crust temperature: (i) they create a local temperature minimum in the crust toward which heat is conducted from both above and below and (ii) they prevent shallow heating from being diffused to higher densities. In the case of MAXI, the effects of (ii) most significantly impact the light curve shape. In order for the crust to be hot enough to match the observations to date the shallow heating must be diffused into the deeper crust where $\rho \gtrsim 10^{11}\ {\rm g \ cm^{-3}}$. However, Urca shells located below $\rho \lesssim 2\times 10^{11}\ {\rm g \ cm^{-3}}$ prevent a large portion of the shallow heat from being diffused deeper, making the crust too cool to agree with late-time observations around $\approx 100\textrm{--}1000$ days (see Figure~\ref{fig:Urca}).  An added complication is that the light curve behavior around $t \sim 1000$ days is degenerate in the model parameters, in particular, the gravity, core temperature, and degree of Urca cooling.

\section{Discussion} 
\label{sec:discussion}

In this study, we fit the quiescent light curve of \maxi \ using crust thermal relaxation models. This source is the hottest quasi-persistent transient yet observed; the fits require a crust temperature $T_b \simeq 2 \times 10^{9} \, {\rm K}$. For this reason, the thermal time in the shallow outer crust is about a factor of 2 shorter than the thermal time in cooler sources like KS~1731-260 and MXB~1659-29 \citep{brown09}. The crust temperature profile is initially steep when entering quiescence, and when combined with a short thermal time early in quiescence, leads to a short exponential decay time when compared to other sources, as noted by \citet{homan14}. The thermal evolution in the solid crust is largely independent of our choice for \qimp\ because the high crust temperature keeps the crust in a regime where electron-phonon scattering dominates over electron-impurity scattering ($T \gtrsim \Theta_{\rm D}$). The high temperature makes the $T_{\rm eff}\textrm{--}T_{b}$ relation largely independent of the light element mass in the envelope. The crust is hot enough to balance crust heating with Urca cooling (i.e., $T_b >{2}\times10^{8}\,{\K}$) and Urca cooling sources impact the light curve shape when located $\lesssim 2\times 10^{11}\ {\rm g \ cm^{-3}}$ (or within a factor of 20 in density of the shallow heat source). For this reason, the presence of Urca cooling pairs in MAXI is disfavored by the observations.

The 16 month outburst is sufficiently long to heat the crust out of thermal equilibrium, but the crust does not reach steady-state before the end of the outburst --- \xte \ also did not reach steady-state during its 19 month outburst \citep{page2013}. Despite the similar outburst accretion rate and outburst duration, MAXI enters quiescence much hotter than \xte. Furthermore, \xte \ requires no shallow heating \citep{page2013} and this would suggest a fundamental difference between the two sources that allows the shallow heating mechanism to operate in MAXI and not in \xte. Surprisingly, however, MAXI does not have large shallow heating during its $2$ month reheating event --- similar to the $\approx 2$ month outburst seen in Swift~J174805.3-244637 where the crust cooling can be fit without shallow heating \citep{degenaar2015}. Interestingly, shallow heating is needed after a similar $\approx 2.5$ month outburst in IGR~J17480-2446 \citep{degenaar2011IGR, degenaar2011mnras, degenaar2013}.

MAXI requires more shallow heating than other sources; the $\approx$ $4\textrm{--}10\,\MeV$ per accreted nucleon of shallow heating is larger than required in KS~1731-260 and MXB~1659-29, each requiring $\approx 1\, \MeV$ per accreted nucleon \citep{brown09}. This hints at an energy source much larger than the $\approx 0.2\, \MeV$ per accreted nucleon supplied by compositionally driven convection in the ocean \citep{medin2011,medin2014} or the $\approx 2\, \MeV$ per accreted nucleon additional deep crustal heating possible given the uncertainties on the nuclear symmetry energy \citep{steiner2012}. The Keplerian energy of the accretion flow is $\sim 80\, \MeV$ per accreted nucleon (at the inner-most stable circular orbit) and may plausibly provide the shallow heating. As suggested by \citet{inogamov99, inogamov2010}, gravitational modes excited in a differentially rotating envelope may dissipate energy deeper in the star. The mode energies are of the order required and the dissipation of these modes in the shallow crust is worthy of future study with realistic ocean and crust models.

The high accretion rate during outburst, when combined with the large amount of shallow heating, brings the crust into a regime of stable helium burning for helium layers at $y \approx  2 \times 10^{8}\, {\rm g \ cm^{-2}}$ \citep{bildsten97, zamfir2014}. During outburst, the crust also enters a regime of stable carbon burning for carbon layers at $y \gtrsim  10^{10}\, {\rm g \ cm^{-2}}$ \citep{cumming2001}. For this reason, an appreciable layer of carbon can not accumulate at the superburst ignition depth around $y \sim 10^{12}\, {\rm g \ cm^{-2}}$. There have been no type-I X-ray bursts or superbursts observed from MAXI to date, consistent with stable burning of helium and carbon. We predict that MAXI is unlikely to have either type-I X-ray bursts or superbursts if strong shallow heating occurs during subsequent accretion outbursts.

\acknowledgements

A.D. thanks Zach Meisel and Hendrik Schatz for useful discussions. Support for A.D. and E.F.B. was provided by the National Aeronautics and Space Administration through Chandra Award Number TM5-16003X issued by the Chandra X-ray Observatory Center, which is operated by the Smithsonian Astrophysical Observatory for and on behalf of the National Aeronautics Space Administration under contract NAS8-03060. A.C. is supported by an NSERC Discovery grant, is a member of the Centre de Recherche en Astrophysique du Qu\'ebec (CRAQ), and an Associate of the CIFAR Cosmology and Gravity program. D. P. is partially supported by a grant from Mexican Conacyt (CB-2009-01, No. 132400). The authors are grateful for support received as part of the International Team on Neutron Star Crusts by the International Space Science Institute in Bern, Switzerland. This material is based upon work supported by the National Science Foundation under Grant No. PHY-1430152 (JINA Center for the Evolution of the Elements).

\bibliographystyle{apj}

\bibliography{paper}

\end{document}